\documentclass[prx,twocolumn,aps,epsf,showpacs,superscriptaddress,longbibliography,nofootinbib]{revtex4-1}
\usepackage[pdftex]{graphicx}
\usepackage[normalem]{ulem}
\usepackage{verbatim, float, complexity}
\usepackage{dcolumn, amsfonts, mathtools}
\usepackage{bm}
\usepackage{latexsym} 
\usepackage{diagbox} 
\usepackage{amsmath}
\usepackage{dsfont}
\usepackage{MnSymbol}
\usepackage[dvipsnames]{xcolor}
\usepackage{array}
\usepackage{bbm}
\usepackage{color}
\usepackage{array}
\usepackage{cancel}
\usepackage{braket}
\usepackage{comment}
\usepackage{titlesec}
\titleformat{\paragraph}[runin]{\bfseries}{}{4pt}{}[:]
\titlespacing{\paragraph}{-4pt}{6pt}{4pt}

\usepackage[colorlinks=true,allcolors=blue]{hyperref}%

\newcommand{\<}{\langle}

\renewcommand{\vec}[1]{{\boldsymbol #1}}
\renewcommand{\>}{\rangle}

\renewcommand{\)}{\right)}

\begin{document}

\title{Observing quantum measurement collapse as a learnability phase transition}

\author{Utkarsh Agrawal}
\affiliation{Kavli Institute for Theoretical Physics, University of California, Santa Barbara, CA 93106, USA}

 \author{Javier Lopez-Piqueres}
 \affiliation{Department of Physics, University of Massachusetts, Amherst, MA 01003, USA}

\author{Romain Vasseur}
\affiliation{Department of Physics, University of Massachusetts, Amherst, MA 01003, USA}

 \author{Sarang Gopalakrishnan}
\affiliation{Department of Electrical and Computer Engineering, Princeton University, Princeton, NJ 08544, USA}

\author{Andrew C. Potter}
\affiliation{Department of Physics and Astronomy, and Quantum Matter Institute, University of British Columbia, Vancouver, BC, Canada V6T 1Z1}

\begin{abstract}
The mechanism by which an effective macroscopic description of quantum measurement in terms of discrete, probabilistic ``collapse" events emerges from the reversible microscopic dynamics remains an enduring open question. Emerging quantum computers offer a promising platform to explore how measurement processes evolve across a range of system sizes while retaining coherence. 
Here, we report the experimental observation of evidence for an observable-sharpening measurement-induced phase transition in a chain of trapped ions in Quantinuum's system model H1-1 quantum processor. This transition manifests as a sharp, concomitant change in both the quantum uncertainty of an observable and the amount of information an observer can (in principle) learn from the measurement record, upon increasing the strength of measurements. We leverage insights from statistical mechanical models and machine learning to design efficiently-computable algorithms to observe this transition (without non-scalable post-selection on measurement outcomes) and to mitigate the effects on errors in noisy hardware.

\end{abstract}
\maketitle

The apparent conflict between the deterministic evolution of wave functions and the probabilistic nature of measurement outcomes has preoccupied physicists and philosophers since the inception of quantum mechanics. The so-called measurement problem entails elucidating the precise mechanism by which a fuzzy quantum state with an uncertain value of an observable, $O$, ``collapses” into one with a sharply determined $O$ value. The crux of this issue lies in explaining how measurement collapse can emerge as an effective description as we scale from the microscopic realm of individual particles and atoms to the macroscopic every-day scale of measurement apparatuses and conscious observers. Quantum computers’ unparalleled capacity to manipulate quantum states of increasingly-large scale while retaining precise and programmable microscopic control offers unprecedented access to explore the evolution of quantum measurement across different system scales.

Inspired by these capabilities, theoretical investigations have revealed striking critical phenomena, known as measurement-induced phase transitions, (MIPTs)~\cite{Li_2018,Skinner_2019,Chan_2019,Choi_2020,Gullans_2020,PhysRevB.101.060301, PhysRevB.101.104301, PhysRevB.101.104302, Potter_2022,Fisher_2023}, that differentiate quantum measurement at microscopic and macroscopic scales. In microscopic quantum systems, such as individual atoms, measurement collapse is a blurry concept, much like the lack of distinction between a solid and a gas in a sample with only a few atoms. Namely, there is smooth crossover between weak-measurements that obtain partial information about an observable while barely perturbing the quantum state, to strong, projective measurements in which uncertainty about the measurement outcome is removed by collapsing the quantum state with real measurements occurring on a spectrum between these two idealized limits. By contrast, in the thermodynamic limit (large system sizes, and long time of interacting with the measurement apparatus), the strong and weak measurement regimes bifurcate into sharply distinct phases separated by a sharp phase transition at a critical measurement strength.

MIPTs arise in a broad class of ``monitored” quantum dynamics that interleave unitary internal evolution of a many-body system with variable-strength interaction with a measurement apparatus. A single experimental run, or ``trajectory”~\cite{Wiseman_1996,breuer2002theory}, produces a (random) measurement record, $M$, and the associated quantum state $|\psi_M\>$.
Early characterizations for MIPTs were phrased in terms of the change in the entanglement structure of the state~\cite{Li_2018,Chan_2019,Skinner_2019} or in the statistical uncertainty in measured-observables~\cite{Sang_2021,morralyepes2023detecting,li2021robust,Bao_2021}, $\<\delta O^2\>_M = \<\psi_M|O^2|\psi_M\>-\<\psi_M|O|\psi_M\>^2$, for each individual trajectory. These characterizations face a fundamentally-insurmountable obstacle to experimental observation: each requires statistical sampling of a large number of copies of a single trajectory.
However, as the measurement record, $M$, is inherently random, repeated experimental runs are extremely unlikely to produce the same post-measured state twice, adding exponential (in the space-time volume of the dynamics) sampling overhead, such that these order parameters are fundamentally unscalable.

\begin{figure*}

\includegraphics[width=1.0\textwidth]{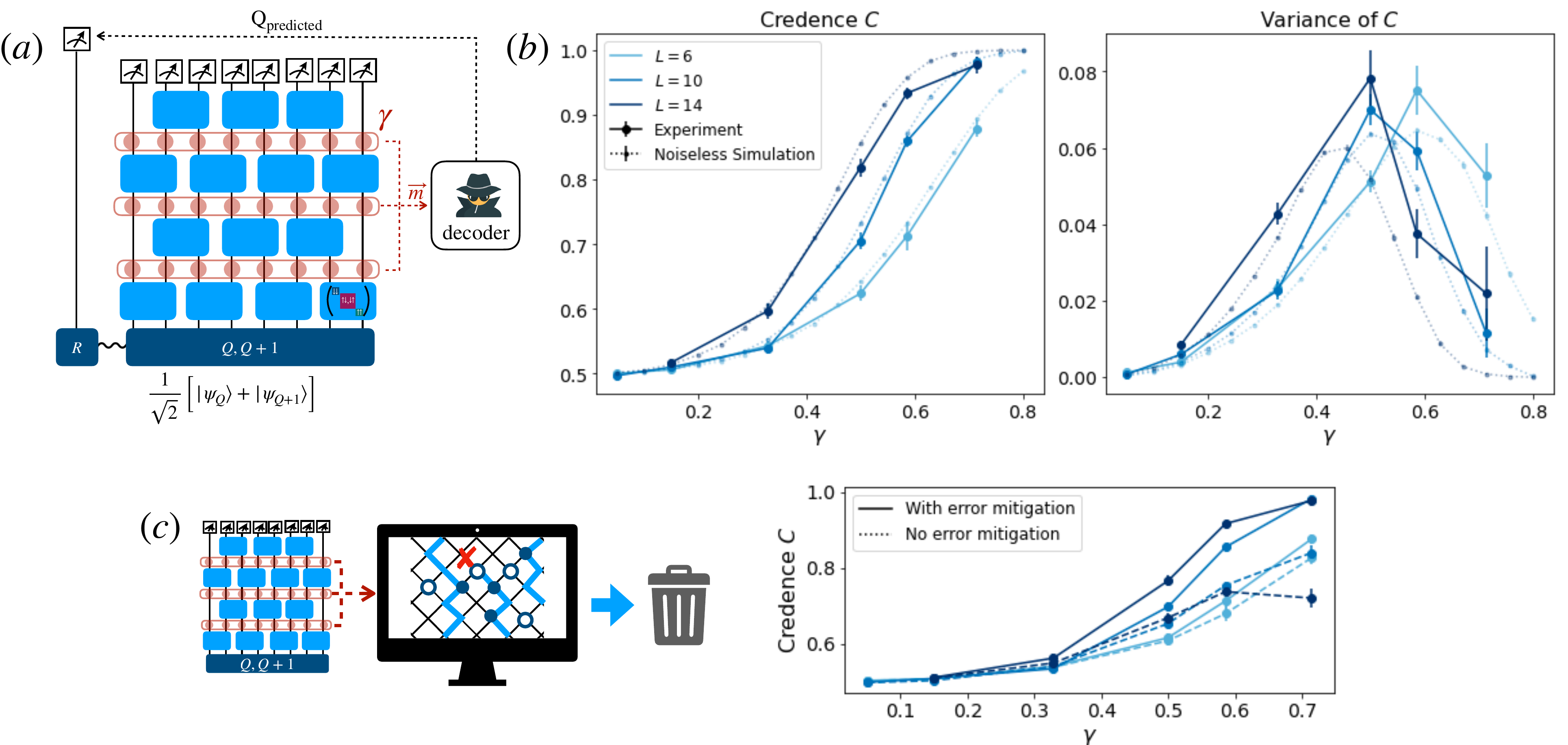}
\caption{{\bf Quantum collapse by learning.} 
(a) Schematic of the monitored quantum circuit and charge-learning protocol.  
A quantum superposition of charge states $Q=L/2$ or $Q=L/2-1$, entangled with a reference, $R$, is fed into a quantum circuit consisting of a brickwork of random charge-conserving gates (blue squares), interspersed with weak measurements of strength $\gamma$ (red dots).
An observer, Eve, processes the measurement data using a decoding algorithm, and attempts to predict the charge. Eve's prediction is then tested against the actual charge, obtained by measuring the reference.
In the experiment, we implement an equivalent protocol, in which we randomly sample initial states $|\psi_Q\>$ or $|\psi_{Q+1}\>$ with equal probability.
(b) Finite-size scaling of Eve's credence $C$ -- the inferred probability that Eve assigns to the correct charge label based on the scalable stat-mech decoding algorithm, averaged over $300-700$ measurement trajectories obtained from experiments using Quantinuum's system model H1-1 trapped-ion quantum processor. 
For low measurement rate, $\gamma$, Eve is unsure of the correct data outcome (average credence well below $1$), and her credence does not improve significantly with system size. For larger measurement rates, $\gamma\gtrsim 0.4$, Eve's performance improves towards $100\%$ as $L$ is increased. 
The variance of $C$ exhibits a peak, which sharpens upon increasing $L$, and converges towards the theoretically-predicted critical measurement strength {$\gamma_c\approx 0.4$}, providing finite-size scaling evidence of a sharp phase transition between the weak and strong measurement regimes.
(c) Data shown in (a,b) uses an error mitigation strategy in which mid-circuit measurements data is used to process and reject samples with charge non-conserving errors that are heralded by the stat-mech decoder assigning $0$ credence to both charge values, $Q,Q+1$.
The percentage of trajectories retained (not discarded by error mitigation) exhibits a weak $\gamma$ dependence, and for the largest measurement strength are: 86\%, 61\%, and 33\%  for $L=6,10,14$ respectively. We note that these error-mitigation overheads are \emph{much} milder than that required to non-scalable methods of observing MIPTs that post-select on measurement trajectories, which would retain only a tiny fraction $2^{-\gamma L^2/2}$ of data ($\approx 10^{-12}$ for $L=14$, $\gamma=0.4$).
} \label{Fig: main fig}
\end{figure*}

In this work, we adopt an alternative learnability perspective on the transition introduced in~\cite{learnability_Fergus} and generalized in~\cite{ippoliti2023learnability}: that evades this post-selection problem, and enables a truly-scalable observation of an MIPT in a generic class of monitored quantum dynamics, which cannot be efficiently simulated by a classical computer. This learnability perspective asks: does an observer, Eve, obtain enough information from the measurement record to accurately predict the value of an observable $O$? 
Successful prediction requires that Eve perform a decoding computation to predict the most-likely outcome of $O$ constant with the measurement record. In a suitably-defined thermodynamic limit, Ref.~\cite{learnability_Fergus} showed that there is a phase transition tuned by measurement strength separating a weak measurement-regime where Eve learns no information from monitoring the system and cannot do better than randomly guessing the value of $O$, and a strong-measurement regime where Eve can infer the value of $O$ with asymptotically-perfect accuracy. Moreover, when Eve employs an optimal decoding strategy, this transition precisely coincides with one in which the uncertainty $\<\delta O^2\>_m$ collapses from its initial value to zero. In contrast to statistical properties of a quantum trajectory, Eve’s accuracy can be checked for each trajectory individually by comparing their prediction against a strong-measurement of $O$ at the end of the dynamics: avoiding the post-selection problem.
For this setting, it is essential that the dynamics of the system and measurement apparatus not scramble the meaning of $O$, for example an initial state state with definite value of $O$ should retain that value for all times.
Measurements satisfying this requirement are dubbed quantum non-demolition (QND) measurements in the quantum optics literature~\cite{braginsky1980quantum}.

Here, we consider simplest type of QND measurement: measurement of an operator $O$ that is conserved by the monitored circuit dynamics.
Following~\cite{learnability_Fergus}, we experimentally examine the task of measuring the total charge, $Q=\sum_i Q_i$ of a chain of $L$ trapped-ion qubits in Quantinuum's system model H1 quantum processor, where the charge of a qubit $i$ is defined as $Q_i = \frac12 (1+Z_i)$, and takes values $0,1$ for the computational basis state $|0,1\>$ respectively.
We engineer a generic (chaotic, non-integrable) $Q$-conserving quantum dynamics by implementing a brickwork of symmetric: two-qubit gates $U = \{u_i\}$
\begin{align}
    u_i =   
    \begin{pmatrix}
    e^{i\phi_0} & &&\\
    &e^{i(\phi_1+\phi_2)}\cos\theta & e^{i(\phi_1-\phi_2)}\sin\theta\\
    &-e^{i(\phi_2-\phi_1)}\sin\theta & e^{-i(\phi_1+\phi_2)}\cos\theta&\\
    &&&e^{i\phi_3}
    \end{pmatrix}
    \label{eq:u}
\end{align}
with parameters $\phi_{0\dots 3}\in [0,2\pi)$, and $\theta \in [0,\pi)$ drawn from the measure: $P(U) = \prod_i\sin\theta_i$ (resulting in Haar-random $2\times 2$ block). Here, we consider a fixed instance of gates, $U$, throughout the experiments.
These unitary gates are interspersed with tunable-strength, $\gamma$, weak measurements, implemented using an ancilla qubit.

The initial state is equiprobably chosen to have fixed charge, either $Q_{\rm correct}=L/2$ or $L/2-1$, unknown to Eve. A ``quantum shuffling" stage hides the initial charge from individual local measurements. This initial state is formally identical to a pure quantum superposition of the two $Q$ values, entangled with a reference qubit.
Eve then observes the measurement record $M$ for $t=L/2$ brickwork layers, and processes $M$ using a decoding algorithm (discussed below) to predict the most likely value of $Q$. Eve's prediction is then checked against the true charge value, $Q_{\rm correct}$.

\begin{figure*}
    \centering
    \includegraphics[width=1.0\textwidth]{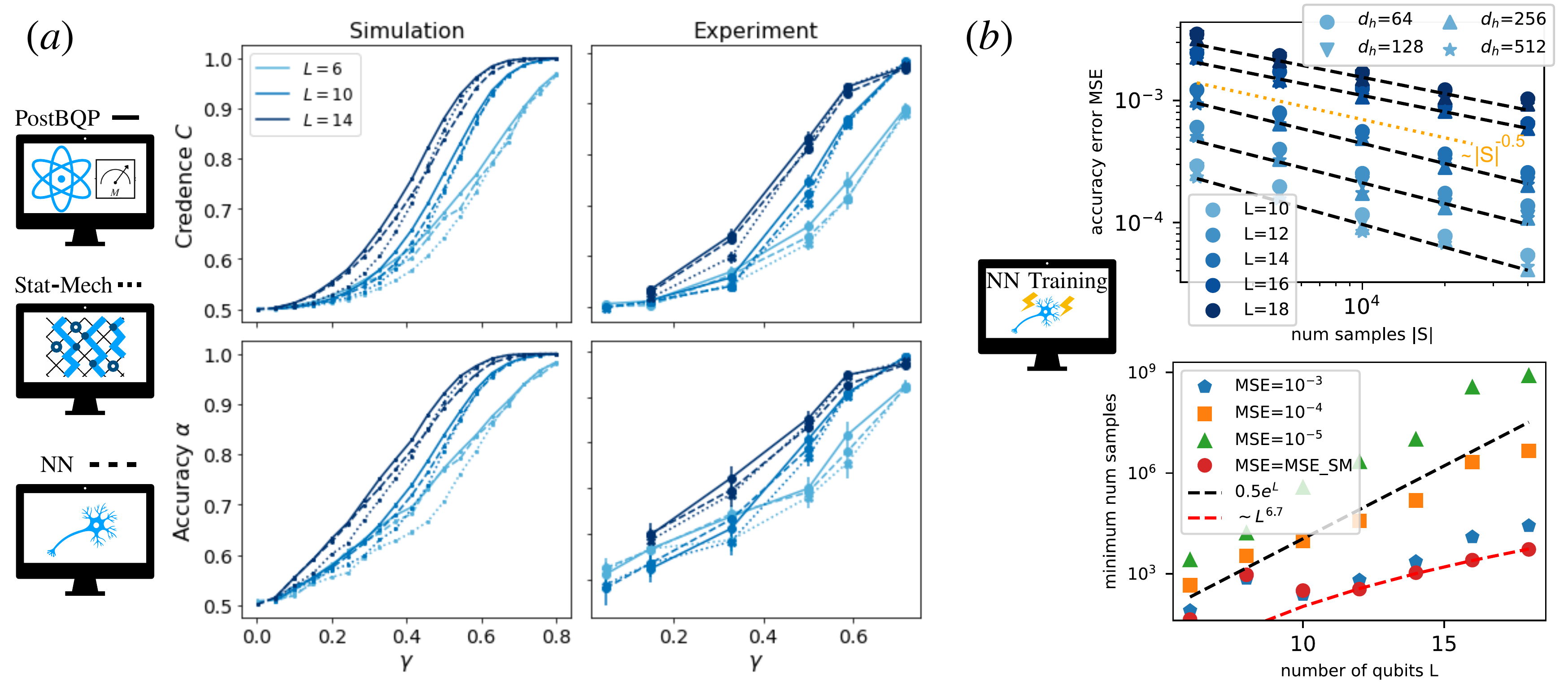}
    \caption{{\bf Decoders and error mitigation.} (a) Comparison of the three decoding algorithms on simulated (noiseless) and experimental data. The simulations of PostBQP and stat-mech decoder are averaged over 10000 trajectories for all $L$ and initial charge $Q$. The RNN is trained over $8000,128000,512000$ samples for $L=6,10,14$ and tested over $2000,32000,128000$ trajectories respectively. This highlights the higher sample complexity for RNN decoder which needs to be trained on exponentially more samples. (b) Scaling behavior of the resources required to perform supervised training of the neural network decoder, on simulated data from a model with strong projective measurements. The data is averaged over different probabilities of measurements. 
    The mean-squared (accuracy) error (MSE) of NN is defined as the mean of the square of the difference between the accuracy of NN and the optimal (PostBQP) decoder. 
    The minimum number of samples required to match the d accuracy of the optimal, PostBQP decoders appears to scale exponentially with system size, $L$ whereas to match that of the classically-efficient stat-mech decoder appears to sub-exponentially (consistent with a power-law dependence) in $L$.
    }
    \label{fig:compare_decoder}
\end{figure*}

\paragraph{Decoders}
Unlike conventional physical phase transitions, which depend only on the dynamics of the system, observable-learning transitions are computational phase transitions that depend jointly on the physical dynamics and the decoding algorithm (``decoder") employed. We consider three different decoders:
\begin{enumerate}
    \item A \emph{PostBQP} decoder based on exact simulation of the quantum dynamics post-selected on the observed measurement outcomes $M$. This decoding algorithm is optimal, but lies in the computational complexity class of post-selected bounded-error quantum polynomial (PostBQP), believed~\cite{aaronson2004quantum} to require exponential (in $L,t$) resources to implement (even on a quantum computer!), and is therefore useful only for validation at intermediate $L$.
    \item A \emph{Statistical Mechanics (stat-mech)} decoder, based on the mapping of the monitored circuit dynamics to a classical stat-mech model~\cite{Agrawal_2022} of random hardcore walkers moving in a potential set by $M$. This decoder can be efficiently implemented by classical matrix-product state methods~\cite{FieldTheory_Fergus}, and is optimal for the case where Eve is ignorant of the quantum phases $\vec{\phi}_i$ of the gates~\cite{learnability_Fergus}.
    \item A recurrent \emph{Neural Network (NN)} decoder trained on a labeled data set which may potentially learn quantum coherent features of the circuit that are ignored by the stat-mech decoder. 
\end{enumerate}

The output of the decoding algorithm is an estimated probability, $D(Q|M,U)$ of the likelihood that the charge was $Q$, conditioned on knowledge of measurements, $M$, and the circuit gates, $U$. Eve then assigns a prediction $Q_{\rm predicted} = {\rm argmax}_{Q}D(Q|M,U)$. To assess whether Eve is successful, we examine both Eve's accuracy: $\alpha_M = 1$ if $Q_{\rm predicted}=Q_{\rm correct}$ and $\alpha_M=0$ otherwise, and Eve's ``credence": the likelihood that Eve assigns to the correct label: $C_M=D(Q_{\rm correct}|M,U)$. We then perform statistical averages of the accuracy and confidence over many ($\sim 300$) measurement rounds, $M$, for fixed set of gates $U$.

The expectation from theory is that, there will be a phase transition separating a strong-measurement regime $(\gamma>\gamma_c)$ where both $\alpha$ and $C$ tend to $1$ as $L\rightarrow \infty$, from a weak-measurement regime $(\gamma<\gamma_c)$ where Eve remains uncertain about the value of the observable ($\alpha,C<1$).~\footnote{Note that the approach to the thermodynamic limit depends in a subtle manner on the order of limits in sending $t,L\rightarrow \infty$~\cite{FieldTheory_Fergus}.}
The critical measurement strength will depend on the decoder used, with the optimal decoding threshold, $\gamma_{c,{\rm PostBQP}}$, lower-bounding that for other decoders. We note that, while the stat-mech decoder exhibits a higher threshold than the PostBQP decoder, the resulting MIPT lies in the same universality class (a modified Kosterlitz-Thouless transition~\cite{FieldTheory_Fergus}), and exhibit identical scaling properties.

\paragraph{Error-Mitigation}
Since the true thermodynamic limit is a theorists' idealization, experimental evidence for a phase transition necessarily requires examining the finite-size scaling with an increasing sequence of system sizes, $L$, and circuit depths $t$. Here, we consider a fixed ratio of size and depth: $t=L/2$.
%
%

As for all practical quantum algorithms, the largest accessible system size is ultimately limited by gate errors. 
In the charge-learning framing, a single charge non-conserving error can cause Eve to predict the wrong charge.
To suppress the effect of noise in finite-size errors, we employ two forms of symmetry-based post-selection. First, we reject samples for which the final measured charge differs from the initial charge (indicating a charge non-conserving error). Second, we reject samples for which the stat-mech decoder assigns zero probability. The latter method is custom-tailored to the charge-learning problem, and significantly improves the quality of the data (See Fig.~\ref{Fig: main fig}.c). 

In the presence of errors, the relevant measure of ``size" of a quantum circuit is not solely the number of qubits, $N$ (where $d$ is the number of spatial dimensions, here $d=1$), but rather, the circuit's entanglement volume ${\rm EV}\equiv \min(N^{1/d},t)^{d+1}$. This quantity represents the volume of space-time regions that can be fully entangled by gates, and is also of direct relevance for the circuit's classical simulation complexity.
Here, the low gate errors in Quantinuum's H1 processor, our error mitigation scheme enables us to achieve circuit volumes of up to $N_{\rm EV}\approx 10^2$, comparable to that of previous MIPT experiments~\cite{google_2023} based on shallow-depth 2d circuits on a nominally much-larger number (70) of qubits.

\paragraph{Finite-size scaling evidence for an MIPT}
To begin, we first examine the behavior of the efficiently-scalable stat-mech decoder, on experimentally-generated quantum data. 
Fig.~\ref{Fig: main fig} shows the statistical estimates of the average and variance of Eve's credence, $C$, based on the stat-mech decoder predictions, for $N_S\sim 300-700$ different experimental shots, with an increasing sequence of system sizes $L=6,10,14$.
The finite-size scaling of these quantities is consistent with the emergence of a sharp phase transition in the large-$L$ limit. Namely, the average $C$ shows two regimes separated: at low measurement rates, $C$ exhibits a weak $L$ dependence, and remains significantly less than $100\%$, at larger measurement rates, $C$ increases with $L$ towards $100\%$. The variance of $C$ peaks at intermediate values of measurement strength, $\gamma$. With increasing $L$ the peak sharpens and the peak location converges towards $\gamma\approx 0.4$, resulting in a crossing of the curves with $L$ -- a hallmark of finite size scaling towards a thermodynamic phase transition.
These results are consistent with the locations of the critical measurement strength $\gamma_{c,{\rm PostBQP}} \approx 0.4 \approx \gamma_{c,{\rm stat-mech}}$ for the optimal and stat-mech decoders determined by classical simulations (see Supplemental Information).

\paragraph{Comparison of decoders}
Fig.~\ref{fig:compare_decoder} shows a comparison of of the credence of the different decoders for experimentally-generated quantum data from the largest system size, $L=14$.
For all measurement rates, the optimal PostBQP decoder outperforms the others. The NN decoder performance lies intermediate between that of the stat-mech and PostBQP decoders, indicating that the NN can successfully learn some phase-coherent features of the particular circuit. 

To assess the efficiency of the NN decoder, we need to consider much larger data sets than are currently feasible to generate with the experimental hardware. To this end, we perform a systematic numerical study of the supervised training of the NN decoder on a large data set generated by classical simulation data of a related monitored circuit model~\cite{Agrawal_2022} with strong projective measurements of a fraction of $p$ qubits, replacing the weak measurements of every qubit. We define the mean-squared error (MSE) of the accuracy as: the average square difference in accuracy: ${\rm MSE} = \<\left(\alpha_{\rm NN}-\alpha_{\rm PostBQP}\right)^2\>$ where $\<\dots\>$ denotes average over gates, and measurement locations and outcomes. We observe the following asymptotic trends.  Training the NN to match the classically-efficient stat-mech decoder appears to require training sets that scale as a polynomial in system size. Indicating that the NN decoder can efficiently match the performance of the classical one.
By contrast, training the NN to closely match (e.g. with MSE$\lesssim 10^{-4}$) the performance of the optimal PostBQP requires exponentially large training sets: $|S|\sim e^{L}$, indicating that it is hard for the NN to accurately learn features related to phase-coherent quantum dynamics that distinguish the PostBQP and stat-mech decoders. For all system sizes, $L$, the MSE decreases as approximately $1/\sqrt{|S|}$ of the number of training data samples, $|S|$, consistent with broadly-valid generalization error estimates of supervised machine learning models.


\paragraph{Discussion and outlook}
The stat-mech insights into monitored random circuits enables both efficiently-computable decoding for the charge-learning problem, and enhanced error-detection capabilities.
Together, these features enabled the experimental observation of finite-size scaling evidence for the existence of an observable-sharpening MIPT. 
In contrast to previous experimental demonstrations of MIPTs, the charge learning approach does not require postselection on measurement outcomes (at exponential in system size overhead)~\cite{koh2023measurement,google_2023}, and also works for generic (e.g. non-Clifford) gate sets that cannot be efficiently simulated classically~\cite{Noel_2022}.
We note that, while the charge-sharpening transition in our qubit-only model occurs at measurement rates where typical trajectories are area-law entangled (and hence can be efficiently simulated by matrix-product techniques), other modified versions of this model~\cite{Agrawal_2022} exhibit the charge-sharpening transition in the highly-entangled regime where classical simulations are not possible.
This raises the intriguing possibility of efficiently observing a MIPT in a classically non-simulable regime. 


These developments suggest several natural avenues for further inquiry: 
Experimentally examining observable-learning MIPTs in 2d qubit arrays, would explore a regime where direct classical simulations quickly become infeasible even for very modest system sizes, yet the stat-mech decoder can likely still be efficiently implemented through Monte Carlo sampling (since the stat-mech model does not have a sign problem). Higher-dimensional circuits also offer qualitatively new features, such as MIPTs at finite-time~\cite{bao2022finite}, and are expected to display unconventional critical phenomena~\cite{FieldTheory_Fergus} without conformal invariance.
Moreover, while the stat-mech decoder transition of the qubit model considered here, occurs in the area-law entangled phase of the circuit dynamics in other closely-related models~\cite{Agrawal_2022,FieldTheory_Fergus} the stat-mech decoder can efficiently and accurately predict the charge in the scrambling, volume-law entangled phase of the circuit dynamics that is believed to be hard to classically simulate.
Can the existence of efficient classical decoders in such classically-non-simulable regimes enable classical verification of quantum advantage? (contrasting existing ``quantum supremacy" protocols based on random circuit sampling~\cite{48651} that paradoxically require classical computations to attempt to show that a quantum algorithm cannot be simulated classically).

Finally, We note that there are multiple existing proposals for evading the post-selection problem to observe MIPTs~\cite{Dehghani_2023,niroula2023phase,Li_2023,garratt2023probing,ippoliti2023learnability}. All fundamentally require cross-correlating experimental quantum data with a computational agent (``decoder") that analyzes the experimental measurement record.
For example, Ref.~\cite{garratt2023probing} proposed examining cross-correlations between quantum (Q) data, and classical (C) simulations of the quantum circuit. 
For this purpose the efficiently-implementable stat-mech decoder may provide a means to scalably observe MIPTs via Q/C correlators in regimes where PostBQP simulations are intractable.
More generally, determining what types of MIPTs can be scalably observed using efficient classical (or quantum) decoding algorithms, and establishing the fundamental computational complexity limits on how closely efficient decoders can approximate optimal performance remain open fundamental questions.


\vspace{6pt}\noindent{\it Acknowledgements -- } 
We thank the entire Quantinuum team for enabling the experimental work, and especially Michael Foss-Feig and David Hayes for insightful discussions.
This work was  supported by the Air Force Office of Scientific Research under Grant No. FA9550-21-1-0123 (R.V. for work on learning transitions); by the US Department of Energy, Office of Science, Basic Energy Sciences, under awards No. DE-SC0019168 and DE-SC0023999 (J.L. for the machine learning component of this work), and DOE DE-SC0022102 (A.C.P.); by the U.S. Department of Energy, Office of Science, National Quantum Information Science Research Centers, Co-design Center for Quantum Advantage (C2QA) under Contract No. DE-SC0012704 (S.G. for conceptualizing the experimental protocol); and by Sloan Research Fellowships (R.V., A.C.P.). We thank the Kavli Institute of Theoretical Physics (KITP) and the Aspen Center for Physics where part of this work were completed. KITP is supported in part by the National Science Foundation under Grant No. NSF PHY-1748958.
This research used resources of the Oak Ridge Leadership Computing Facility, which is a DOE Office of Science User Facility supported under Contract DE-AC05-00OR22725.

\appendix
\section{Methods}
\subsection{Trapped-ion implementation}
We implement the charge-conserving monitored dynamics on Quantinuum’s H1-1 trapped-ion quantum processor~\cite{Pino_2021} which supports up to 20 $^{171}{\rm Yb}^+$ trapped ion qubits, in a Quantum Charge Coupled Device (QCCD) architecture. 
For the largest system size $L=14$, we used $18$ qubits with 4 ancilla qubits used for performing weak measurements. 
The native gates include arbitrary single-qubit (1q), and a native two-qubit (2q) M\o lmer-S\o rensen entangling gate with unitary $e^{i\frac\pi 4 Z\otimes Z}$. At the time of experiments, typical gate errors for 1- and 2-qubit gates determined by randomized benchmarking protocols~\cite{Knill_2008} were respectively: $p_{1q}\approx 4\times 10^{-5}$ and $p_{2q}\approx 2\times 10^{-3}$.

The QCCD architecture enables high fidelity, low cross-talk, mid-circuit measurements and qubit resets by 
Further, the ability to physically-transport ions without effecting their quantum state enables one to perform arbitrary long-range gates between any pair of qubits.
As explained below, we utilize these mid-circuit measurement and qubit re-use and long-range gates to reduce the number of ancilla qubits required, and to effectively prepare pre-scrambled initial states where the total charge is hidden from local measurements.

\paragraph{Initial state preparation}
To prepare the initial state with charge $L/2$, we initialize the qubits alternatively in $\ket{0}$ and $\ket{1}$ (we flip the last qubit in $\ket{1}$ to $\ket{0}$ to get charge $L/2-1$). We then delocalize/scramble the charge to prevent a single round of measurements from learning significant information about the charge. For this, we apply all-to-all gates where we randomly pick $L/2$ pairs and apply 2-qubit U(1) gates to them. We apply $5$ layers of the all-to-all gates. The pairs in each layer are randomly chosen but are fixed for all trajectories. The long-range gate capability of Quantinuum's ion-trap hardware allows us to physically move the qubits around and implement the above all-to-all circuit. After preparing the state using the above prescription we start the monitored dynamics.

\paragraph{Weak measurements}
To perform weak measurements we entangle system qubits with individual ancillas (initialized in $\ket{0}$ state). We took the entangling gate to be a controlled rotation along x-direction $R_x(\gamma \pi/2)$ with the ancilla being the target. Let the state of the system before measurement be $\ket{\psi}=\ket{\psi_0}+\ket{\psi_1}$, where $\ket{\psi_i}$ are unnormalized states and $i$ is the charge of the qubit to be measured. After the controlled rotation the combined state of ancilla and system is given by\begin{align*}
	\ket{\psi}_{SA} = \ket{0}_A\ket{\psi_0} + \(i\sin(\gamma \pi/2)\ket{1}_A + \cos(\gamma\pi/2)\ket{0}_A\)\ket{\psi_1}.
\end{align*}
Projective measurement on the ancilla then implements weak measurement of the charge of the qubit. For the ancilla outcome $1$, the qubit is projected to have a charge whereas for outcome $0$ the probability of having no charge at site $i$ increases but is crucially still less than $1$. For $\gamma=0$ no measurements are being performed and $\gamma=1$ is the projective measurement limit. By tuning the value of $\gamma$ we can tune the strength of weak measurements.
In the QCCD architecture, a single ancilla can be re-used for performing weak-measurements on multiple different physical qubits by physically transporting the ancilla by measuring the ancilla ion, resetting it, and physically transporting it to the location of the next qubit to be measured. 

To implement measurement of all qubits at each time step we find it convenient to assign a single ancilla qubit for measurement of a few system qubits. After each measurement, the ancilla is restored to $\ket{0}$ and used again for measurement of the next qubit. E.g for simulating $L=14$ on the hardware we actually used $18$ qubits where the remaining $4$ qubits served as ancillas.

\subsection{Decoders}
\paragraph{PostBQP} The optimal PostBQP decoder is the exact classical simulation of the circuit dynamics post-selected on the observed measurement outcomes. To be more precise we calculate $D_\mathrm{PostBQP}(Q|M,U)$ as follows. We initialize the quantum state used in the experiment with charge $Q$ and run the quantum circuit dynamics with fixed measurement outcomes $M$ on a classical computer. We calculate the Born probabilities $P(M|Q,U)$ for obtaining measurement outcomes $M$ for the initial charge is $Q$. We run the above simulation for all $Q$ used in the experiment to get $P(M|Q,U)$. $D_\mathrm{PostBQP}(Q|M,U)$ is then given by $P(M|Q,U)/(\sum_q P(M|q,U))$ where the sum over $q$ is over all initial charges used in the experiment. Though optimal, the above decoder requires exponential-in-$L$ resources and cannot be scalably implemented with either a classical or quantum (due to post-selection) processor. We refer to this decoder as a PostBQP decoder because the complexity of the decoder lies in PostBQP complexity class~\cite{aaronson2004quantum} consisting of problems that are solvable in a hypothetical quantum computer with polynomial (in $L$) circuit depths, and post-selection on measurement outcomes. PostBQP is believed to contain problems that cannot be simulated efficiently by a real (non post-selected) quantum computer. To our knowledge, the implementation of this classical decoder would require exponential in system-size and circuit depth resources to carry out on either a classical or quantum device.

\paragraph{Stat-mech}
 We next consider an approximate decoder obtained by marginalizing over the phases of the random gates in (\ref{eq:u}). For each run with outcome $M$ we define the marginalized probability $P'(M|QU)=\int d\phi_i P(M|Q,U)$ where $P(M|Q,U)$ is the Born probability to observe outcomes $M$ in the exact quantum evolution, and $\phi_i$ are the quantum phases in the unitary gates in eq~\eqref{eq:u}. We then define the stat-mech decoder $D_\mathrm{stat-mech}=P'(M|Q,U)/(\sum_q P'(M|q,U))$. Numerically evaluating the above integral is even harder than running the PostBQP decoder. But as shown in~\cite{learnability_Fergus}, the marginalized probabilities, $P'(M|Q,U)=\int d\phi_i P(M|Q,U)$ can be analytically simplified and written as the partition function of a \emph{classical} noisy symmetric exclusion process (SEP). Crucially, this classical model can be efficiently simulated, e.g. by matrix-product state methods~\cite{FieldTheory_Fergus}. However, for the range of system sizes explored in this experiment, it was similarly efficient to directly implement the transfer matrix approach, via standard (non-MPS-based) numerical matrix-vector multiplication routines, to calculate $P'(M|Q,U)$ in the stat-mech model. 

\begin{figure}
    \centering
    \includegraphics[scale=0.36]{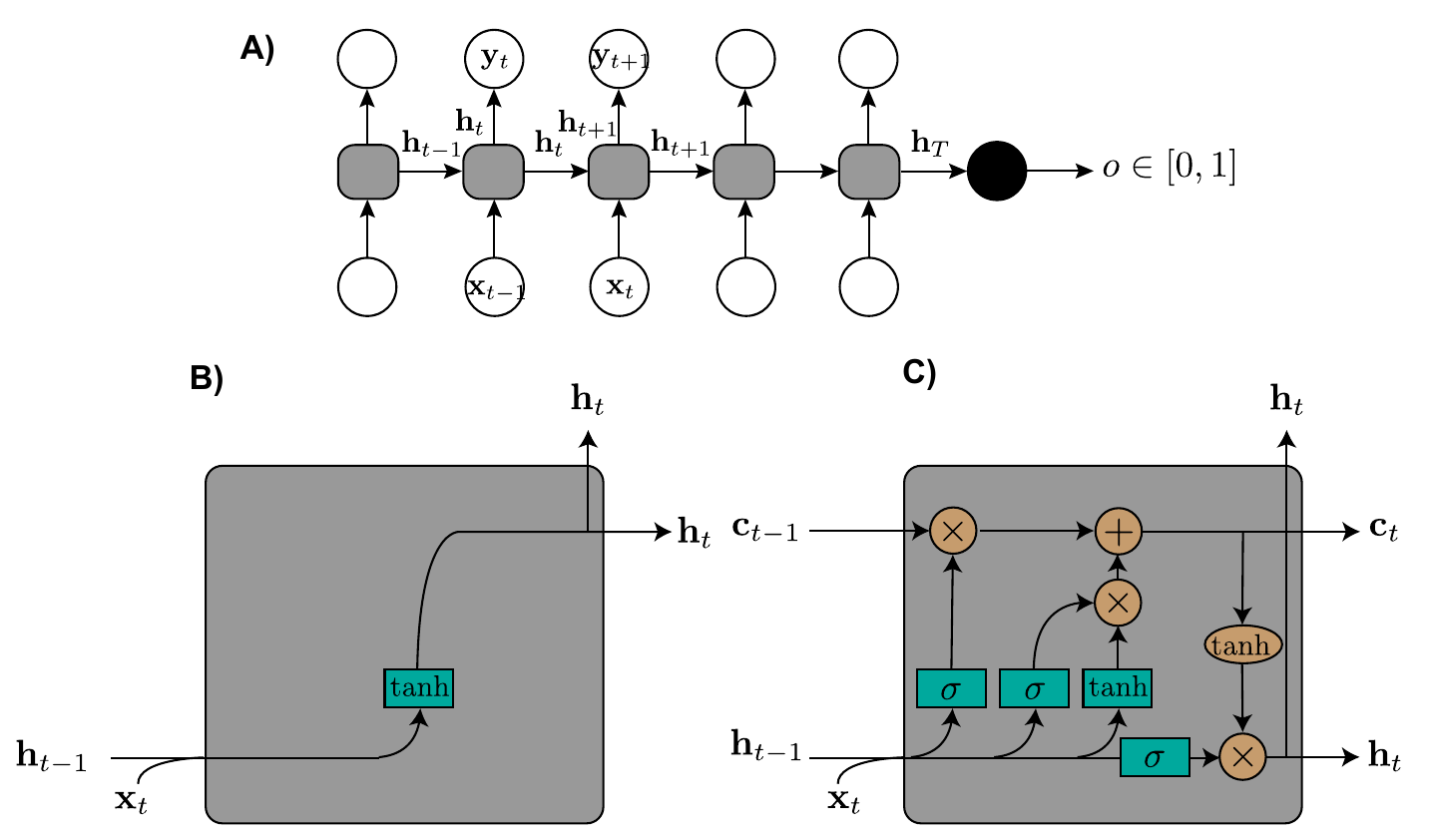}
    \caption{\textbf{RNN architecture.} A) Unrolled RNN diagram. B) Cell of a vanilla RNN. C) Cell of an LSTM. Within each cell, rectangles denote NN layers with tanh or sigmoid activation functions. Circles denote pointwise addition or multiplication. Single arrows denote vector transfers. Merging and forking of two arrows denote concatenation and copying of two vectors, respectively.}
    \label{fig:RNN-diagram}
\end{figure}

\paragraph{Neural Network}
Finally, we consider neural networks (NNs) as classical decoders. The goal of the network is to construct a model of the distribution $P(Q^*|M)$, from a collection of measurement records. In contrast with SEP and PostBQP, this family of decoders require training over a set of measurement records, and the quality of the decoder is evaluated over some testing set. Given the causal, time-series structure of the measurement records we adopt a recurrent neural network (RNN). RNNs exploit the \textit{autoregressive property} of time-series whereby the model distribution up to some time $T$ depends on all previous time steps. More precisely, given a set of measurement records $\{(\vec{x}_1, \vec{x}_2, \cdots, \vec{x}_T)\}$, with $T$ some number of time steps, and denoting $P(\vec{x})\equiv P(\vec{x}_1,\vec{x}_2, \cdots, \vec{x}_T)$ the probability for a given measurement record, the RNN computes this via the factorization property 
\begin{equation}
P(\vec{x})=\prod_{t=1}^T P(\vec{x}_t|\vec{x}_1,\vec{x}_2,\cdots,\vec{x}_{t-1}).
\end{equation} 
The diagram of an RNN is shown in Fig. \ref{fig:RNN-diagram}. The vectors $\{\vec{h}_t\}$ are called \textit{hidden} vectors and the $\{\vec{y}_t\}$ are called \textit{output} vectors. The role of the hidden vector is to encode information about sequences. The dimension of each hidden vector will be the same and of value $d_h$, sometimes referred to as the \textit{number of memory units/cells}, and the bigger the more expressible the network is. The output vector will be of the same dimension as the input vector. The RNN updates its hidden and output vectors according to the following dynamical rule
\begin{align}
\vec{h}_t&=f(W_h\vec{x}_{t-1}+U_h \vec{h}_{t-1}+\vec{a}), \\
\vec{y}_t&=g(U_o\vec{h}_{t-1}+\vec{b}),
\end{align}
where $W$ and $U$ are (weight) matrices and $\vec{a}$, $\vec{b}$ are biases. All these are free parameters that the RNN must update so as to find a good model of $P(\vec{x})$. The functions $f$, $g$ are functions that are taken to be nonlinear in general (often called \textit{nonlinear activation} functions), with common choices being \textit{softmax} and \textit{tanh}. The application of this function is pointwise on each vector entry. At every time step one can compute the joint probability distribution via one-hot encoding the input vectors and performing the scalar product 
\begin{equation}
P(\vec{x}_t|\vec{x}_1,\vec{x}_2,\cdots, \vec{x}_{t-1})=\vec{y}_t \cdot \vec{x}_t,
\end{equation}
so long as there's guarantee that $||\vec{y}_t||=1$, which is the case for nonlinear activation functions with image in $[0,1]$ (such as softmax). 

In this work we are interested in using RNNs for binary classification, where the two labels correspond to the two different charges considered, $Q_0=L/2$ and $Q_1=L/2-1$. To model $P(Q^*|M)$, we collapse the last hidden vector into a number in $[0,1]$ via a softmax. Though the goal of the RNN is to maximize the classification accuracy, we can take as proxy for the conditional probability above the following quantity $P(Q^*|M) \approx 1-|o-q^*|$ where $o$ is the output of the RNN for a given measurement record and $q^*$ is the correct label, being $q^*=0$ if $Q^*=L/2$ and $q^*=1$ if $Q^*=L/2-1$. This proxy naturally reproduces the classification accuracy given as 
\begin{equation}
\alpha=\mathbb{E}[P(Q^*|M)>0.5] \approx \frac{1}{|\mathcal{V}|}\sum_{\vec{x} \in \mathcal{V}}\chi_{|o_{\vec{x}}-q^*_{\vec{x}}| < 0.5},
\end{equation}
where $\mathcal{V}$ is the testing set of measurements and $\chi_*$ is the indicator function. 

While one could in principle optimize the RNN parameters so as to maximize the accuracy, it is often preferred to minimize instead the binary cross-entropy. This is in great part because the latter is less sensitive to class imbalances. It is also faster to train via gradient descent and leads to better generalization when compared to other loss functions, such as sum-of-squares \cite{simard2003best}. In our case, it is even more crucial because a given measurement record may in principle correspond to different labels, and the goal of the RNN is to predict the probability a given measurement record corresponds to a given charge. 

A common problem when training RNNs via gradient descent is that they may lead to vanishing or exploding gradients when capturing long-range correlations between input variables~\cite{chung2014empirical,bengio1994learning}. To prevent this, sophisticated RNN architectures such as the gated recurrent unit (GRU) and long-short term memory (LSTM) NNs have been proposed. Here we will use the LSTM architecture to capture those long distance correlations. The basic cell diagram is shown in Fig. \ref{fig:RNN-diagram}. It contains more nonlinear maps than the vanilla counterparts and includes the presence of a new vector, $\vec{c}_t$, also called \textit{cell state}. It barely has any interaction with the other vectors and only through linear maps. Its whole purpose is to deal precisely with these long term dependencies \cite{45500}.

\textit{Numerical details.} We train an LSTM using the TensorFlow library. To minimize the training cross-entropy while avoiding overfitting we compute the testing accuracy after every epoch and stop training whenever the testing accuracy has plateaued for five or more epochs. For the biggest problem sizes of $L=14-18$ number of qubits and depths of size $L$ and the largest number of memory cells, $d_h=256-512$, only an hour or two of computation is needed when using a single GPU. We use the Adam optimizer~\cite{kingma2014adam} to implement the stochastic gradient updates on the binary cross-entropy.
Empirically, we observe that the NN performance is relatively insensitive to the details of the network, such as the number of cells or type of network (see Supplemental information Fig.~\ref{fig:transformer}).

\begin{figure}[t!]
    \centering
    \includegraphics{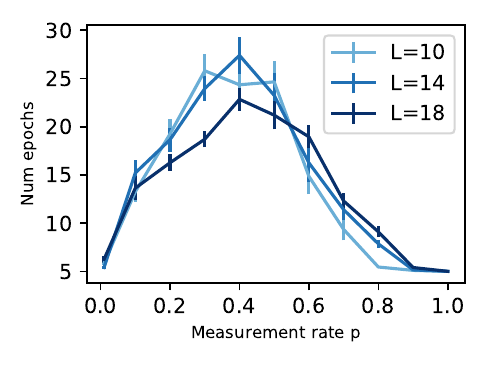}
    \caption{\textbf{Time complexity for probing the transition in NNs}. Number of epochs needed to achieve convergence in accuracy when training RNNs for different circuit sizes. The depth of the circuit is $L$. Error bars correspond to averaging over 20 circuit realizations, and the number of measurement records is fixed to 40000 for each charge.}
    \label{fig:num_epochs}
\end{figure}
\paragraph{Training epochs} In Fig. \ref{fig:num_epochs} we analyze the learning complexity transition from counting the time required, as measured by the number of training epochs, for the RNN to achieve convergence in accuracy when decoding circuits with projective measurements. We stop training whenever the accuracy has plateaued for 5 or more epochs. 
The number of training epochs peaks at a measurement probability near $p\approx 0.4$. 
This behavior is possibly suggestive of a training-complexity phase transition at some measurement rate $p\approx 0.4$. We note that, since the learnability/decoding phase transitions depend jointly on both the system dynamics and the decoding algorithm, this putative transition would not necessarily coincide with those for the the optimal PostBQP or stat-mech decoders that respectively occur at, $p_{c,{\rm PostBQP}}\approx 0.1$, $p_{c,{\rm stat-mech}}\sim 0.2$. However, with the presently-accessible range of system and training data set sizes, we are currently unable to cleanly resolve the evolution of the behavior of the peak height and shape with system size, $L$, and leave a more detailed investigation into this feature for future work. 

\subsection{Error mitigation}
Table.~\ref{table: post-selection} lists the percentage of samples discarded as part of the error mitigation described in the main text. Samples are discarded if either: i) the final measured charge is different than the initial charge, or ii) the stat-mech decoder assigns vanishing credence to the correct label $C_M = D(Q_\mathrm{correct}|M,U)=0$.
The discarded fraction grows significantly with system size $L$, as expected since the dominant error mechanism arises from two qubit entangling gates. The discarded fraction also weakly increases with measurement strength $\gamma$, indicating that stronger mid-circuit measurements are better able to detect (and discard) configurations with errors.

\begin{table}[b!]
\begin{tabular}{|c||c|c|c|c|c|c|}
    \hline
    \diagbox[width=3em]{$L=$}{$\gamma\approx$} & 0.05 & 0.15 & 0.33 & 0.5 & 0.58 & 0.71 \\
    \hline
    \hline
    &&&&&&\\[-0.8em]
    6 & 11.5\% & 9.2\% & 7.5\% & 8.5\% & 18.5\% & 14.5\%\\
    \hline
    &&&&&&\\[-0.8em]
    10& 34.5\% & 36.5\% & 38.2\% & 34.5\%  & 41\% & 38.7\% \\
    \hline
    &&&&&&\\[-0.8em]
    14& - & 57.1\% & 61.7\% & 65.9\% & 65.4\% & 66.5\% \\
    \hline 
\end{tabular}
\caption{{\bf Error mitigation statistics} Percentage of samples discarded for each system size, $L$, and measurement strength, $\gamma$.}\label{table: post-selection}
\end{table}

\bibliography{u1mrc}

\appendix
\section*{Supplementary Information}

\begin{figure*}
    \centering
    \includegraphics[width=0.7\textwidth]{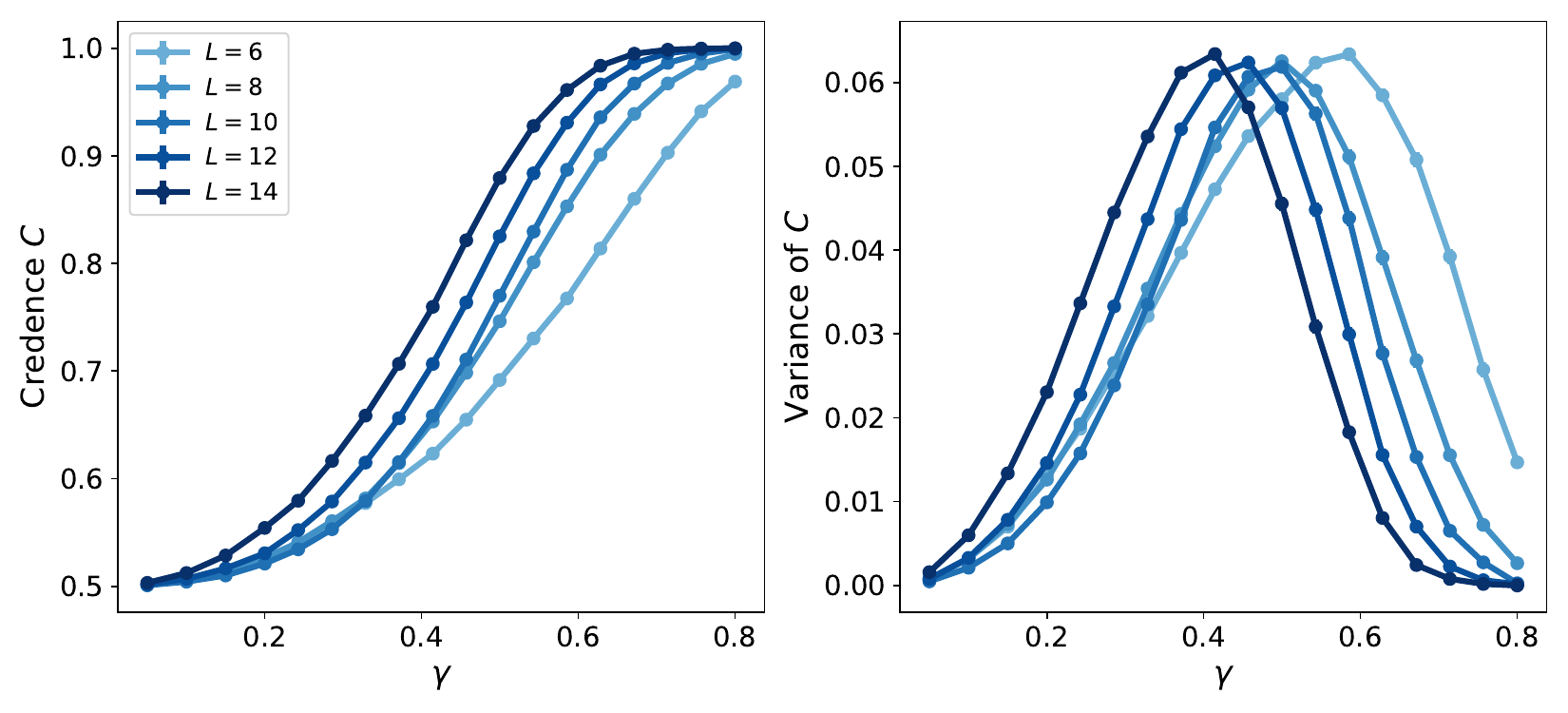}
    \caption{\textbf{PostBQP decoder transition.} Plots for credence $C$ and its variance on simulated data for PostBQP decoder. Number of samples is equal to $10000$.}\label{fig: postbqp simulation}
\end{figure*}
\begin{figure*}
    \centering
    \includegraphics[width=0.7\textwidth]{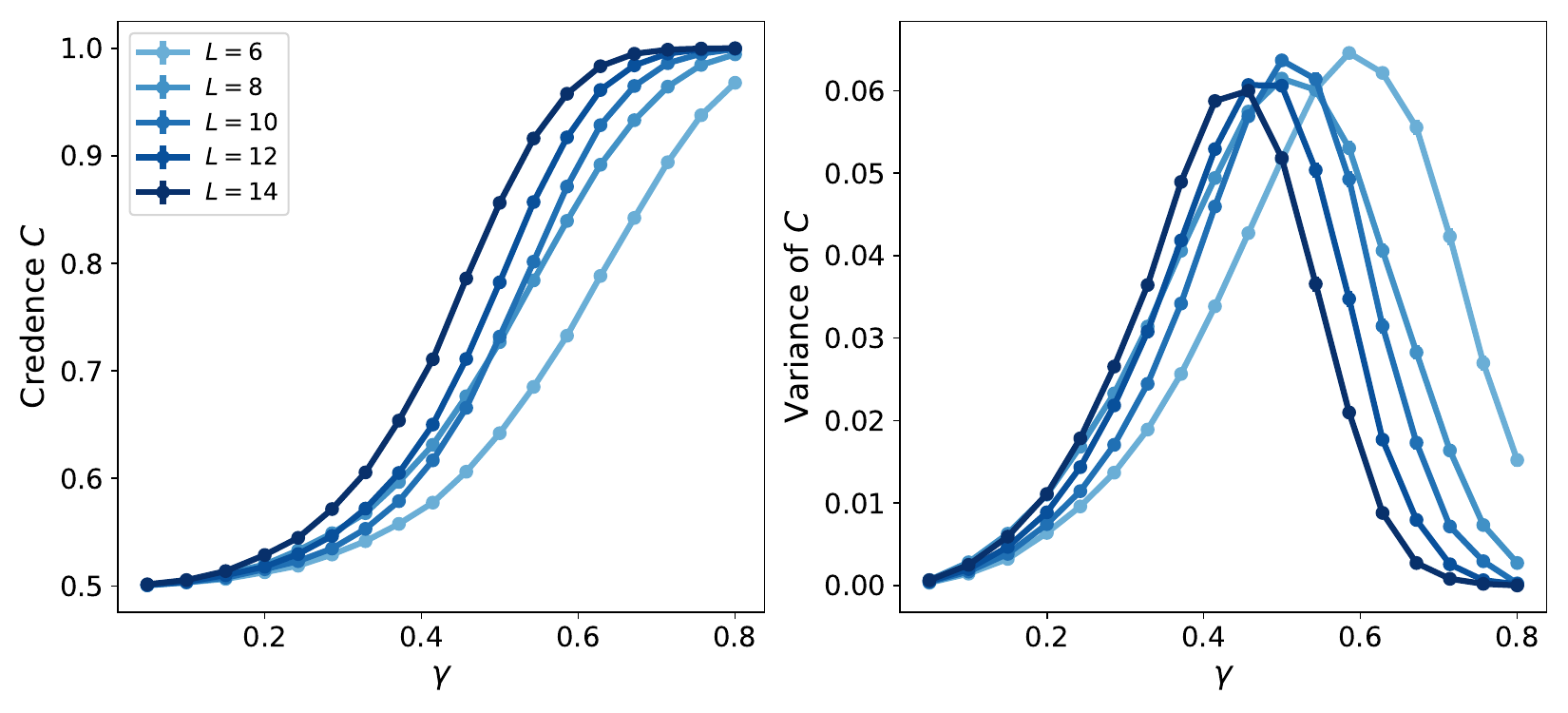}
    \caption{\textbf{Stat-mech decoder transition.} Plots for credence $C$ and its variance on simulated data for the stat-mech decoder. Number of samples is equal to $10000$.}\label{fig: stat-mech simulation}
\end{figure*}

\section{Stat-mech (SEP) decoder}
The stat-mech decoder described in the main text has a nice interpretation: it is the optimal decoder Eve can use in case she only has access to the classical information about the charge transport, or due to reasons best known to her, she is not able to efficiently use the quantum knowledge of the dynamics. This decoder has been argued to have a sharp charge-learning phase transition in the thermodynamic-limit, at a higher critical measurement strength $\gamma_{c,{\rm SEP}}\geq \gamma_{c,{\rm opt}}$, than the optimal decoder, but with precisely the same universal scaling properties.

Here we describe the details about the stat-mech decoder we used for the experiments. The decoder is a classical dynamical model for the charge degrees of freedom. The model has been described in detail in~\cite{learnability_Fergus} the only difference being instead of projective measurements we perform weak measurements. Let us first note how projective measurements were implemented before modifying them for the weak measurements. At any point in time, the classical state of the system is a probability distribution over all possible local charge configurations, $P(\{q_i\};t)$. If a projective measurement with outcome $m_i$ was made at time $t$ and qubit $i$, the probability distribution is changed to $P(q_i\neq m_i ;t) \rightarrow 0$ while $P(q_i =m_i ;t)$ remains unchanged (except for the overall normalization factor).

For weak measurements, depending on the measurement outcome of the ancilla, we have the following update rules,
\[
\begin{aligned}
    &P(q_i = 0 ;t) \rightarrow 0 && \mathrm{outcome}=1 \\
    \ \\
    &\begin{aligned}
	P(q_i = 0 ;t) &\rightarrow P(q_i = 0 ;t) \\
	P(q_i = 1; t)  &\rightarrow \cos^2(\gamma\pi/2)P(q_i = 1 ;t)
\end{aligned} && \mathrm{outcome}=0
\end{aligned}
\]
with, of course, the distribution being normalized later on.

\begin{figure*}
    \centering
    \includegraphics[width=0.6\textwidth]{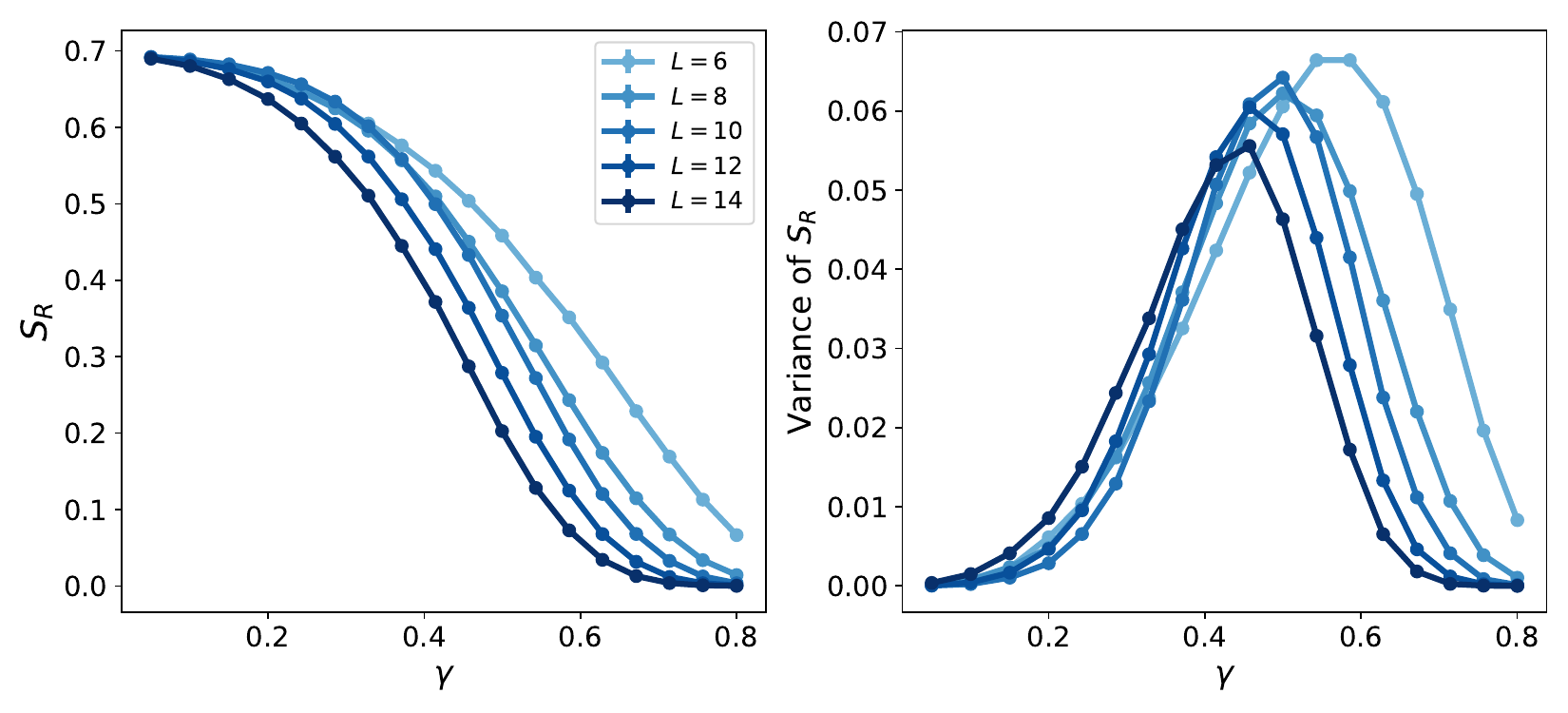}
    \includegraphics[width=0.32\textwidth]{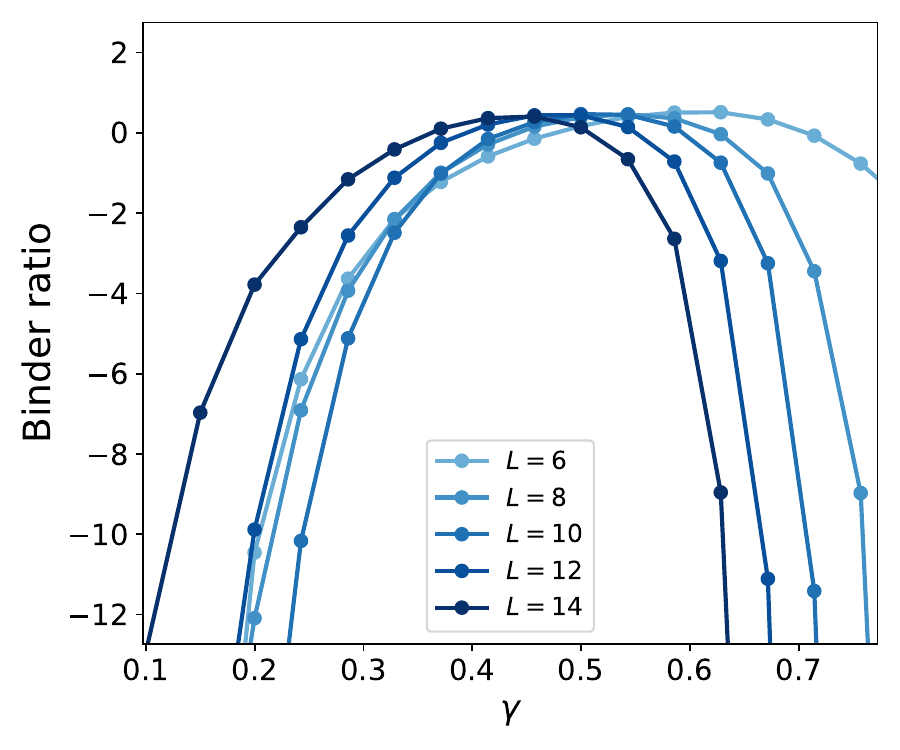}
    \caption{\textbf{Charge sharpening.} \textit{Left.} Plot for ancilla entropy in the charge sharpening setup described in eq \ref{eq: charge sharpening}. Number of samples is equal to $10000$. \textit{Center.} Variance of $S_R$.     \textit{Right.} Binder ratio for the ancilla entropy defined in eq \ref{eq: binder}. We can see a crossing near the critical point.}\label{fig: ancilla data}
\end{figure*}

\subsection{Optimality}
The stat-mech decoder described in the main text has a nice interpretation: it is the optimal decoder Eve can use in case she only has access to the classical information about the charge transport, or due to reasons best known to her, she is not able to efficiently use the quantum knowledge of the dynamics. This decoder has been argued to have a sharp charge-learning phase transition in the thermodynamic-limit, at a higher critical measurement strength $\gamma_{c,{\rm SEP}}\geq \gamma_{c,{\rm opt}}$, than the optimal decoder, but with precisely the same universal scaling properties.

Here we prove the optimality of the stat-mech decoder in the absence of knowledge of the unitary gates. The optimal PostBQP decoder requires Eve to know and use the full unitary dynamics. If for some reason Eve doesn't have access to the gates she can optimize over the unitary gates to get the best possible circuit\begin{align*}
    U_\mathrm{opt} = \mathrm{argmax}_U \alpha[U],
\end{align*}  
where $\alpha[U]$ is the trajectory average of the accuracy for circuit $U$. However, this is computationally costly and also infeasible to find the optimal in the large search space of the unitary gates.

We can instead think of $U$ as a statistical model to classify a given measurement trajectory $M$ to one of the possible charges. Then it is a well-known result in machine learning~\cite{madigan1994model,hoeting1999bayesian} that the optimal classifier in such cases is one where the models are averaged. That is, the best Eve can do is\begin{align}
    D(Q|M) = \int dU D_{\rm PostBQP}(Q|M,U) P(M|U),
\end{align}
where $P(M|U)$ is the probability to get outcomes $M$ given model $U$. Typically averaging over the models is a hard task but in our case, we can average over $U$ analytically to get a stat-mech model for $D(Q|M)$~\cite{learnability_Fergus}. We can also choose the average over partial information about $U$, for example, the phases in the gates as considered in the main text. 

\section{Simulation data}
\subsection{Learnability transition}
We plot the credence $C$ and its variance for the PostBQP decoder simulation with $10000$ samples in Fig.~\ref{fig: postbqp simulation}. We find the critical point (located using the peak of the variance) to be drifting to lower values and saturating around $\gamma_c \lesssim 0.4$.

We also plot the credence $C$ and its variance for the stat-mech decoder simulation with $10000$ samples in Fig.~\ref{fig: stat-mech simulation}. We find the critical point (located using the peak of the variance) again drifting to lower values but saturating at a higher strength around $\gamma_c \lesssim 0.45$.

\subsection{Charge sharpening transition}
Following~\cite{Agrawal_2022}, we can study charge sharpening transition by introducing an ancilla $R$ and entangle it with the system in the state,\begin{align}
    \ket{\psi}_{SR} = \frac{\ket{Q_0}_S \ket{0}_R + \ket{Q_1}_S\ket{1}_R}{\sqrt{2}}, \label{eq: charge sharpening}
\end{align}
where the subscript $S,R$ denote the system and the ancilla respectively. The charge sharpening transition can be detected by looking at the entanglement entropy of $R$, $S_R$, at late times. We plot $S_R$ vs measurement strength $\gamma$ for the circuit used to run the experiment in FIg.~\ref{fig: ancilla data}. We expect the charge sharpening transition to be the same as the charge-learnability transition for the optimal PostBQP decoder. We also define the binder ratio $B$ of the ancilla entropy \begin{align}
    \mu_4 = &\sum_m p_m (S_{R}^{(m)} - S_R)^4, \nonumber \\
    \mu_2 = &\sum_m p_m (S_{R}^{(m)} - S_R)^2, \nonumber \\
    B = &\frac{1-\mu_4}{3\mu_2^2}, \label{eq: binder}\\ 
\end{align} 
where $S_R^{(m)}$ is the entropy of $R$ for trajectory $m$, $S_R = \sum_m p_m S_R^{(m)}$ is the average entropy. We plot $B$ in Fig.~\ref{fig: ancilla data} and see a drifting crossing near $\gamma_c\approx 0.4$.

\section{Additional data for Neural Network training}
This appendix contains additional data comparing the performance of different neural network architectures.

\begin{figure}[b!]
    \centering
    \includegraphics[scale=1]{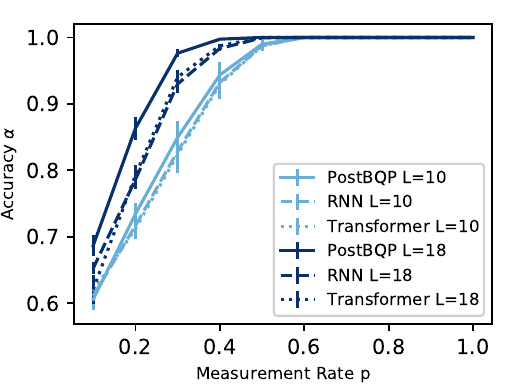}
    \caption{\textbf{NN decoder benchmarks.} Decoding performed over 10 circuits and 40000 measurement records for each charge and for each circuit. Training is performed on 80\% of the samples and tested for accuracy on the remaining. The depth of the circuits are $L$.}
    \label{fig:transformer}
\end{figure}

 We have also tried other neural network architectures with the aim of testing the robustness of our learning complexity claims. In particular, we have tried the transformer architecture \cite{vaswani2023attention}, which is one of the state-of-the-art architectures for sequence modeling. When compared to LSTMs, the transformer has been shown to handle longer range correlations better due to its self-attention mechanism \cite{shen2019mutual}. In our case, since we are doing binary classification we only resort to the encoder part of the architecture containing an embedding layer followed by transformer block which includes self-attention and feed-forward layers. Our results show that overall the performance of the LSTM is on par with the transformer. In Fig. \ref{fig:transformer} we benchmark the performance of fine-tuned transformers vs fine-tuned LSTMs when decoding circuits with projective measurements. The results are then averaged over 10 circuits, where each circuit contains 40000 measurement records for each charge (80000 in total) and each NN is trained over 80\% of the measurement records (and tested on the remaining 20\%). The code used to reproduce these results can be found on Ref. \cite{github-repo}.

\end{document}